\documentclass[aps,12pt,superscriptaddress,tightenlines]{revtex4}
\usepackage{amsmath,amssymb,amsthm,graphicx,bm,color}

\newcommand{\p}{\partial} 
\newcommand{\supp}{\textrm{supp\,}}

\renewcommand{\Delta}{\varDelta} 
\renewcommand{\Gamma}{\varGamma} 
\renewcommand{\Omega}{\varOmega} 
\renewcommand{\Phi}{\varPhi} 
\renewcommand{\Psi}{\varPsi} 
\renewcommand{\Sigma}{\varSigma} 
\renewcommand{\Theta}{\varTheta} 
\renewcommand{\epsilon}{\varepsilon}

\newcommand{\R}{\mathbb{R}}
\newcommand{\Z}{\mathbb{Z}}

\theoremstyle{definition}
\theoremstyle{definition}

\definecolor{myred}{rgb}{1,0,0}

\begin{document}

\title{Higher order topological invariants from the Chern-Simons
action}

\author{Roman V. Buniy}

\email{roman@uoregon.edu}

\affiliation{Institute of Theoretical Science, University of Oregon,
Eugene, OR 94703}

\author{Thomas W. Kephart}

\email{tom.kephart@gmail.com}

\affiliation{Department of Physics and Astronomy, Vanderbilt
University, Nashville, TN 37235} 

\date{\today}

\begin{abstract}
  It is well known that for a field theory with the Chern-Simons
  action, expectation values of Wilson line operators are topological
  invariants. The standard result is expressed in terms of the
  Gaussian linkings of closed curves defining the operators. We show
  how judicious choice of Wilson lines leads to higher order
  topological linkings.

\end{abstract}


\maketitle

\section{Introduction}

Our understanding of knots and links in topological field theory began
with the work of Polyakov \cite{Polyakov} who showed that the
expectation value of a Wilson line in the Chern-Simons theory gives
the Gaussian linking of the components of a link. This work was soon
followed by Witten's \cite{Witten} extensive exploration of
expectation values of Wilson lines in the non-abelian Chern-Simons
theory which he showed is a natural framework for understanding Jones
polynomials from the knot theory. Perturbative expansion of the
expectation values leads to the third order linking and the
corresponding link polynomials~\cite{GMM}. Our approach is somewhat
different. Wilson loops in the abelian Chern-Simons theory also
implicitly contains the requisite apparatus for the study of higher
order linking, and we show here that this can be facilitated by a set
of gauges \cite{BK1,BK2} tailored to the higher order linking
problem. We begin with a discussion of intersection number and its
relation to Gaussian linking. Next we review the minimal required
background from Poincar\'{e} duality and the de Rham theorem needed to
carry out our calculations. We then introduce a special set of gauge
potentials that allow us to arrive at our main result--higher order
linking invariants at all orders.

\section{Intersection and linking numbers}

Suppose $C$ and $C'$ are disjoint oriented closed curves in $\R^3$,
and $S$ and $S'$ are surfaces such that $\p S=C$ and $\p S'=C'$. By
smooth deformations of $C$ which leave it in $\R^3-C'$, we can change
the number of points of intersection of $C$ and $S'$. To find a
quantity which is invariant under such deformations, we note that
additional points come in pairs, and intersections of $C$ and $S'$
have opposite orientations for points in each pair. If $C$ and $S'$
intersect transversely at a point $p$, we define the intersection
index $I(C,S',p)$ to be equal $1$ or $-1$ depending on the relative
orientation of $C$ and $S'$ at $p$. We also define the intersection
number $I(C,S')$ as the sum of the intersection indices over all
points of the intersection,
\begin{align}
  I(C,S')=\sum_{p\in C\cap S'} I(C,S',p).
\end{align}
It is clear that $I(C,S')$ is invariant under the above deformations
as contributions due to additional pairs of points cancel.

One can also deform $S'$ into $\tilde S'$ and notice that
$I(C,S')=I(C,\tilde S')$, where $\p S'=\p \tilde S'=C'$. This means
that the intersection number depends only on $C$ and $C'$; for this
reason we call it the linking number, $L(C,C')=I(C,S')$. Examination
of relevant orientations gives $L(C,C')=L(C',C)$. The method of 
Green functions leads to an explicit expression for the linking number
of the curves in terms of their parametrizations,
\begin{align}
  L(C,C')=(4\pi)^{-1}\int_{C\times
  C'}\sum_{abc}\epsilon_{abc}\frac{(x-x')^a}{\|x-x'\|^3}dx^b\wedge
  {dx'}^c.
\end{align}

\section{Duality\label{section:duality}}

From the Poincar\'{e} duality and de Rham
theorem~\cite{algebraic-geometry,topology}, for a closed curve $C$ in
three dimensions, there exists a closed $2$-form $F$ such that for any
$1$-form $B$ we have
\begin{align}
  \int_C B=\int_{\R^3} B\wedge F.
\end{align}
We call $(C,F)$ a dual set. Since $B$ is arbitrary, it is clear that
$\supp F=C$. Since $\R^3$ is simply connected, there exists a $1$-form
$A$ such that $F=dA$. The Stokes theorem then
gives~\cite{stokes-theorem}
\begin{align}
  \int_S dB=\int_{\R^3} dB\wedge A.
\end{align}
This means that there is a particular solution $A$ such that $\supp
A=S$,
\begin{align}
  A(x)=\int_{y\in S}\delta(x-y)\sum_a dx^a *dy^a.
\end{align}
We can similarly introduce a dual set $(C',F')$ such that $F'=dA'$ and
for any $1$-form $B'$ we have
\begin{align}
  \int_{C'} B'=\int_{\R^3} B'\wedge F'.
\end{align}
Taking $B=A'$, $B'=A$ and using the Stokes theorem again, we find
\begin{align} 
  \int_{C}A' =\int_{C'}A =\int_{\R^3} A\wedge dA' =L(C,C').
\end{align}
Since $L(C,C')$ is a topological invariant, smooth deformations of $C$
which leave it in $\R^3-C'$ should not change its value. This is
possible if and only if $dA'\vert_{\R^3-C'}=0$. If $A'$ is exact, then
$L(C,C')=0$, and so a nontrivial case is when $A'$ is a closed
$1$-form which is not exact.

\section{Second order fields}

Suppose $\{C_i\}_{1\le i\le N}$ are disjoint closed curves in $\R^3$
and let $F_i=dA_i$ be dual to $C_i$. We now construct a dual set
$(C_{ij},F_{ij})$ of the second order which satisfies
\begin{align}
  \int_{C_{ij}} B=\int_{\R^3} B\wedge F_{ij}
\end{align}
for any $1$-form $B$ and $i\not=j$. Since $B$ is arbitrary, it follows
that $\supp{F_{ij}}=C_{ij}$.  The most general $2$-form which can be
expressed in terms of $A_i$ and $A_j$ is
\begin{align}
  F_{ij}=f_i dA_j -f_j dA_i +gA_i\wedge A_j,
\end{align}
where $f_i$, $f_j$ and $g$ are arbitrary functions. In order for
$F_{ij}$ to be closed, the functions have to satisfy certain
conditions. A requirement $dF_{ij}\vert_{\R^3-C_i-C_j}=0$ gives
$dg\vert_{\R^3-C_i-C_j}=0$ and without loss of generality we set
$g=1$. Requirements $dF_{ij}\vert_{C_i}=0$ and $dF_{ij}\vert_{C_j}=0$
then give
\begin{align}
  (df_j-A_j)\vert_{C_i}&=0,\\
  (df_i-A_i)\vert_{C_j}&=0.
\end{align}
Integrating these conditions, we find a constraint
$L(C_i,C_j)=0$. This means that the second order field associated with
a pair of closed curves can be defined only if the curves are
unlinked.

\begin{figure}[ht]
  \includegraphics[width=5cm]{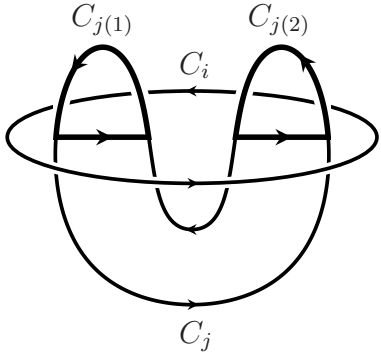}
  \caption{\label{figure-second-order-curve} Two components of
  $C_{ij}$ are drawn with thick lines. $S_i$ is a disk bounded by
  $C_i$, and the thick line segments in this disk are two disjoint
  segments of $S_i\cap S_j$.}
\end{figure}

Since $dF_{ij}=0$, there exists a $1$-form $A_{ij}$ such that
$F_{ij}=dA_{ij}$. We seek a solution in the form
\begin{align}
  A_{ij}=\tfrac{1}{2}(\gamma_i A_j -\gamma_j A_i).
\end{align}
From a requirement $F_{ij}=dA_{ij}$, the unknown functions $\gamma_i$
and $\gamma_j$ are found to satisfy
\begin{align}
  (d\gamma_i-A_i)\vert_{\R^3-C_i-C_j}&=0,\\ d\gamma_i\vert_{C_i}&=0,\\
  (d\gamma_i-2A_i)\vert_{C_j}&=0,\\ (2f_i-\gamma_i)\vert_{C_j}&=0
\end{align}
and the same expressions with $i$ and $j$ interchanged.  This implies
$\gamma_i\vert_{\R^3-\cup_i C_i}=\int_{\Gamma_i} A_i$, where
$\Gamma_i$ is a curve in $\R^3-\cup_i C_i$; this means $\gamma_i$ is a
nonlocal quantity. If $S_i\cap C_j=\varnothing$, then
$d\gamma_i\vert_{C_j}=0$, and so $\gamma_i\vert_{C_j}$ is a
constant. If $S_i\cap C_j\not=\varnothing$, then $S_i\cap S_j
=\cup_m\bigl(S_i\cap S_j\bigr)_{(m)}$, where $m$ labels disjoint
segments of the intersection; for an example, see
Fig.~\ref{figure-second-order-curve}.
Let $C_{j(m)}$ be the segment of $C_j$ which closes the curve
$\bigl(S_i\cap S_j\bigr)_{(m)}$ and agrees with its orientation; this
closed curve is $C_{ij(m)}=C_{j(m)}\cup\bigl(S_i\cap
S_j\bigr)_{(m)}$. We define $C'_j=\cup_m C_{j(m)}$ and its complement
in $C_j$ is $C_j''$. It follows from the above relations that
$\gamma_i\vert_{C'_j}$ and $\gamma_i\vert_{C''_j}$ are constants such
that $\gamma_i\vert_{C'_j}-\gamma_i\vert_{C''_j}=2$. Without loss of
generality, we set $\gamma_i\vert_{C'_j}=2$ and
$\gamma_i\vert_{C''_j}=0$. Using the definition of $F_{ij}$, the
duality condition now becomes
\begin{align}
  \int_{C_{ij}}B=-\int_{C'_i}B +\int_{C'_j}B +\int_{S_i\cap S_j}B.
\end{align}
This implies $C_{ij}={C'_i}^{-1}\cup C'_j \cup\bigl(S_i\cap
S_j\bigr)$, which agrees with $\supp{F_{ij}}=C_{ij}$. See
Fig.~\ref{figure-second-order-curve} for an example of the above
construction. The Stokes theorem gives
\begin{align}
  \int_{S_{ij}} dA=\int_M dA\wedge A_{ij},
\end{align}
where a surface $S_{ij}$ is such that $\p S_{ij}=C_{ij}$. This means
$\supp A_{ij}=S_{ij}$.

The above construction requires $L(C_i,C_j)=0$. Since the curves $C_i$
and $C_j$ are unlinked, the surfaces $S_i$ and $S_j$ can be chosen to
be disjoint, in which case $dA_{ij}\vert_{\R^3-(C_i\cup C_j)}=0$. One
can similarly proceed to construct higher order fields. For example,
$A_{ijk}$ for $i\not=j\not=k$ is required to satisfy
$dA_{ijk}\vert_{\R^3-(C_i\cup C_j\cup C_k)}=0$. If all linkings of
order $q$, where $1\le q\le p\le N-1$ vanish, then we can similarly
construct dual sets $(C_{I_q},dA_{I_q})$, where $I_q=(i_1,\ldots,i_q)$
and $i_1\not=i_2\not=\cdots\not=i_q$. The quantities $dA_{I_q}$ are
related to what is known in algebraic topology as the Massey products
of cohomology groups~\cite{Massey}; see also \cite{Berger}.

\section{Path integral}

We want to construct a field theory for which expectation values of
observables are topological invariants of various orders. For this we
need to specify the action and the observables. We choose the
Chern-Simons action,
\begin{align}
  S(B)=\int_{\R^3}B\wedge dB,
\end{align}
since it has all the necessary topological properties
\cite{Witten}. First, it does not depend on the choice of
metric. Second, it can be related to linking numbers in the following
way. Suppose $\Gamma_\alpha$ is a closed curve on which $dB$ takes a
constant value $dB_\alpha$. (In order for $\Gamma_\alpha$ to be
closed, we may need to identify points at infinity by considering
$S^3$ instead of $\R^3$. We also avoid field configurations with
sources, like in monopoles.) Since a union of curves
$\Gamma=\cup_\alpha\Gamma_\alpha$ densely fills $\R^3$, an arbitrary
$1$-form $B$ can be written as $B=\sum_\alpha B_\alpha$. The action
becomes
\begin{align}
  S(B)=\sum_{\alpha\beta} L(\Gamma_\alpha,\Gamma_\beta),
\end{align}
and so (apart from the choice of measure for $\Gamma$) we interpret it
as the self-linking $L(\Gamma,\Gamma)$ of the set of closed field
lines of $dB$.

If $C=\cup_\alpha C_\alpha$ is a union of disjoint closed curves, then
an integral $\int_C B$ is invariant with respect to deformations of
$C$ if and only if $dB=0$. Since $dB=0$ is the classical equation of
motion for $S(B)$, the integral is a topological observable (at least
in the semiclassical limit). Proceeding as above, we find
\begin{align}
  \int_C B=\sum_{\alpha\beta} L(\Gamma_\alpha,C_\beta),
\end{align}
which we interpret as the linking $L(\Gamma,C)$ of the sets of closed
field lines of $dB$ with $C$. Since the measure in the path integral
is $\exp{(iS(B))}$, it is convenient to consider as an observable a
Wilson loop operator $W(C,B)=\exp{(i\int_C B)}$. 

We thus need to compute the expectation value
\begin{align}
  Z(C) =\int DB \, \exp{(iS(B))}W(C,B).
\end{align}
Using duality, this becomes
\begin{align}
  Z(C) =\int DB \, \exp\biggl(i\int_{\R^3} B\wedge d(B+A)\biggr),
\end{align}
where $dA$ is dual to $C$. Changing the variable $B=B'-\tfrac{1}{2}A$
gives four terms in the exponent. One corresponds to the path integral
of the measure without a Wilson loop, $Z(\varnothing)$, the two mixed
terms combine into a boundary term, and the forth is the linking
invariant. Ignoring the boundary term, we find
\begin{align}
  Z(C) =Z(\varnothing) \exp\bigl(-\tfrac{1}{4}i(L(C,C))\bigr).
\end{align}

If we take $C=\cup_p\cup_{I_p}C_{I_p}^{n_{I_p}}$, where
$C_{I_p}^{n_{I_p}}$ is a curve $C_{I_p}$ repeated $n_{I_p}\in\Z$
times, then we find
\begin{align}
  L(C,C)=\sum_{pq}\sum_{I_pJ_q}n_{I_p}n_{J_q}L(C_{I_p},C_{J_q}).
\end{align}
Here $L(C_{I_p},C_{J_q})$ is the first order linking of curves
$C_{I_p}$ and $C_{J_q}$; it can also be interpreted as a linking of
order $p+q-1$ of the curves $\{C_i\}_{1\le i\le N}$. In general,
$L(C,C)$ is a sum of all linkings of orders $1\le r\le 2N-1$. We now
look at two simple examples.

\section{Examples}

\begin{figure}[ht]
  \includegraphics[width=6cm]{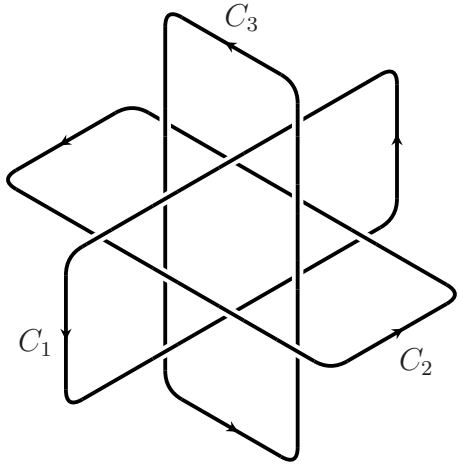}
  \caption{\label{figure-borromean-rings} The Borromean rings.}
\end{figure}

The simplest example is when only the second order linkings appear. We
take $C=\cup_{1\le i\le N}C_i$ and find
\begin{align}
  L(C,C)=\sum_{ij}L(C_i,C_j),
\end{align}
in agreement with a result in Refs.~\cite{Polyakov, Witten}. For the
third order linking, we consider the simplest
example~\cite{rolfsen,kauffman} of the Borromean rings in
Fig.~\ref{figure-borromean-rings}.
 Since $L(C_i,C_j)=0$ for $i\not=j$, we can define the corresponding
$C_{ij}$, and so we take $C=C_1\cup C_2\cup C_3\cup C_{12}\cup
C_{23}\cup C_{31}$. The second order linking vanishes and the third
order linking is given by
$L_3(C,C)=\sum_{ijk}\epsilon_{ijk}L(C_{ij},C_k)$. Comparison of
Figs.~\ref{figure-second-order-curve} and \ref{figure-borromean-rings}
shows that $L(C_{12},C_3)=1$. This can also be seen from
\begin{align}
  L(C_{12},C_3)=\int_{\R^3}A_1\wedge A_2\wedge A_3
  -\tfrac{1}{2}\int_{C_1}\gamma_2 A_3 +\tfrac{1}{2}\int_{C_2}\gamma_1
  A_3
\end{align}
since in the special gauge the first term equals $1$, two
contributions to the second term cancel, and the third term
vanishes. The symmetry then leads to $L_3(C,C)=6$.

\section{Conclusions}

In the Chern-Simons theory, the first order (Gaussian) linking of two
curves $C_1$ and $C_2$ associated with the 1-forms $A_1$ and $A_2$ can
be ascribed to the topological properties of the expectation value
that the associated Wilson lines are found to obey. Here we have
generalized the idea to the case of higher order linking. First, if
$C_1$ and $C_2$ are unlinked, then we are free to define a new second
order 1-form $A_{12}$ via Eq. (11). The associated curve $C_{12}$,
unlike $C_1$ and $C_2$, is not fixed in space, but is however
sufficient for our topological needs, which are to investigate its
linking with other curves. If for instance we have a third curve $C_3$
unlinked with both $C_1$ and $C_2$, but linked with $C_{12}$, then we
find nonzero second order linking, i.e., a nonvanishing second order
topological invariant, and a detailed treatment of the Borromean rings
we given as an example of our method. The argument generalizes to
higher orders. For curves $\{C_i\}_{1\le i\le N}$, with no pairwise
linking, no tertiary linking, etc., only $(N-1)$th order linking, the
general results are obtained from the path integral of the expectation
value of $N$ Wilson lines in the Chern-Simons gauge theory. One could
apply these results to the investigation of higher order link
polynomials.

\begin{acknowledgments}
 The work of RVB was supported by DOE grant number DE-FG06-85ER40224
 and that of TWK by DOE grant number DE-FG05-85ER40226.
\end{acknowledgments}


\begin{thebibliography}{100}
\bibitem{stokes-theorem} We assume that the value of appropriate
   quantities vanishes at infinity, so there is no contribution from
   the boundary term. Boundary terms also vanish if we use $S^3$
   instead of $\R^3$.
\end{thebibliography}
\end{document}